\documentclass[aps,preprint]{revtex4}
\usepackage{amssymb}
\usepackage{amsmath}
\usepackage{graphicx}
\usepackage{subfigure}
\usepackage{epstopdf}

\begin{document}

\title{Thermodynamics of nonlinear charged Lifshitz black branes with
hyperscaling violation}
\author{M. H. Dehghani $^{1,2}$\thanks{%
mhd@shirazu.ac.ir}, A. Sheykhi $^{1,2}$ \thanks{%
asheykhi@shirazu.ac.ir}, and S. E. Sadati $^{1}$ }
\address{$^1$ Physics Department and Biruni Observatory, College of
Sciences, Shiraz University, Shiraz 71454, Iran\\
$^2$ Research Institute for Astronomy and Astrophysics of Maragha
(RIAAM), P.O. Box 55134-441, Maragha, Iran}
\begin{abstract}
In this paper, we investigate the thermodynamics of hyperscaling violating Lifshitz black branes
in the presence of a nonlinear massless
electromagnetic field. We, first, obtain analytic nonlinear charged black
brane solutions with hyperscaling violating factor in dilaton gravity and give the
condition on the parameters of the metric for having black brane solutions.
Second, we introduce the appropriate finite action in grand-canonical and
canonical ensembles for nonlinear electromagnetic field. Next, by
generalizing the counterterm method for the asymptotic Lifshitz spacetimes
with hyperscaling violating factor, we calculate the energy density of our solutions.
Then, we present a relation between the energy density and the
thermodynamic quantities, electric potential, charge density, temperature
and entropy density. This relation is the generalization of Smarr formula
for anti-de Sitter black branes and charged Lifshiz solutions. Finally, we
perform a stability analysis in both the canonical and grand-canonical
ensemble. We show that the nonlinearity of electromagnetic field can make
the solutions unstable in grand-canonical ensemble.
\end{abstract}

\maketitle

\section{Introduction}

Gauge/gravity duality \cite{J. M. Maldacena} may be thought of as a
practical tool to study strongly coupled systems near critical points where
the system exhibits a scaling symmetry. Generally, at a critical point the
system may be described by a conformal field theory (CFT). From the
gauge/gravity point of view this means that the gravitational theory is
defined on a metric which is asymptotically locally anti-de Sitter (AdS). On
the other hand, the central idea in gauge/gravity duality is that each state
in the bulk has a corresponding state in the dual field theory. In
particular, black objects are dual to thermal ensembles in the field theory
with the same thermodynamic properties (temperature, entropy, chemical
potential, etc.) as the bulk spacetime. Most notably, the holographic
techniques are also useful to study condensed matter systems at strong
coupling \cite{Hartnoll}.

Although the idea of holography has been first used in the AdS/CFT
correspondence, largely inspired by condensed matter systems these
techniques have been brought to bear on other types of systems too. As a
first generalization, one may consider metrics which can be dual to
scale-invariant field theories which are, however, not conformally
invariant, but instead enjoy a dynamical critical exponent $z\neq 1$ \cite%
{Bala}:%
\begin{equation}
ds^{2}=-r^{2z}dt^{2}+\frac{dr^{2}}{r^{2}}+r^{2}d\vec{x}%
_{d-2}^{2}.  \label{met1}
\end{equation}%
This metric which is known as Lifshitz metric is invariant under the
following scaling
\begin{equation}
t\rightarrow \lambda ^{z}t,\;\;\;\;\;\;r\rightarrow \lambda ^{-1}r\;\;\;\;\;x^i\rightarrow
\lambda x^{i}.  \label{scale}
\end{equation}%
For $z=1$, the scaling is isotropic; it corresponds to relativistic
invariance. In \cite{Bala}, it was proposed that gravity duals of field theories
with Lifshitz scaling should have metric solutions that asymptote to
Lifshitz metric give in Eq. (\ref{met1}). Indeed, the Lifshitz black holes
have been found to emerge as gravity duals of some condensed matter systems
with anisotropic scaling symmetry (\ref{scale}). It is well known that the
Einstein-Hilbert (EH) action does not admit Lifshitz geometry with or
without hyperscaling violation. In order to have Lifshitz solutions, one may add a
massive gauge field to the EH action \cite{Lif,Deh,Peet1,Peet2}. Another way of constructing
asymptotically Lifshitz solutions is through the use of a dilaton field \cite%
{Dil}. The exact asymptotically Lifshitz charged solutions of
Einstein-dilaton-Maxwell gravity have been introduced in \cite{Tarr}.

Another generalization which has been recently got more attention is the
study of systems with hyperscaling violation. Indeed, by including both
dilaton and Abelian gauge fields, it is possible to find even more
interesting metrics, which on the top of an anisotropic scaling, have also an
overall hyperscaling violating factor. More precisely, one may have a geometry in the
form of \cite{Charm}
\begin{equation}
ds^{2}=r^{\frac{-2\theta }{d}}\left( -r^{2z}dt^{2}+\frac{dr^{2}}{r^{2}}+%
r^{2}d\vec{x}_{d-2}^{2}\right) ,  \label{met2}
\end{equation}%
where the constants $z$ and $\theta $ are dynamical and hyperscaling
violating exponents, respectively. This is the most general geometry which
is spatially homogeneous and covariant under the scale-transformation (\ref%
{scale}) and is not scale invariant, but transforms as
\begin{equation}
ds\rightarrow \lambda ^{\frac{\theta }{d}}ds,
\end{equation}%
under the transormations (\ref{scale}).
The concept of hyperscaling violation has also developed in condensed matter
and in the context of gauge gravity duality \cite{Dong}. For example,
hyperscaling violation provides a useful way to realize compressible matters
in $(2+1)$-dimensions and to pursue the holographic realization of systems
with Fermi surfaces \cite{L. Huijse1}. The EH action with Abelian gauge
field coupled to a dilaton may also have black hole solutions which
asymptote to the hyperscaling violating Lifshitz metric given in Eq. (\ref{met2}). For
asymptotically Lifshitz spacetimes with hyperscaling violation, black
hole solutions were obtained in \cite{Alishahiha:2012}.

Although an exact asymptotically Lifshitz black brane solution with
hyperscaling violating factor has been presented in \cite{Alishahiha:2012}, its
thermodynamics has not been investigated. In the present paper we would like
to investigate the effects of adding a power-law Maxwell invariant term to
the action of Einstein-Maxwell in the presence of a massless scalar dilaton field. This
higher term, $(-F_{\mu \nu }F^{\mu \nu })^{s}$, also appears in the
low-energy limit of heterotic string theory \cite{M. Hassaine}. Also, it is
worth noting that this term for $s=d/4$, is conformally invariant \cite{M.
Hassaine}. We also like to investigate the thermodynamics of hyperscaling violating Lifshitz black
branes in the presence of a nonlinear
massless gauge field. In order to do this, we should generalize the
counterterm method for Lifshitz black holes
in the presence of a massive electromagnetic field introduced in \cite{saremi2009}. We introduce the
appropriate action of this theory in both the canonical and grand-canonical
ensembles. The appropriate action of Einstein gravity both in canonical and
grand-canonical ensembles in the presence of a linear gauge field and in the
absence of dilaton has been introduced by Hawking and Ross \cite{S. W.
Hawking grand}. This action has been generalized in \cite{DSV} for nonlinear
electromagnetic field. Our first aim, here, is to generalize this action for linear
electromagnetic field to the case of nonlinear gauge field both in canonical
and grand-canonical ensembles in the presence of dilaton. Having the
appropriate action in hand, we use the counterterm method to calculate the
finite energy density of the black brane solutions.

This paper is outlined as follows. In the next section we present the action
and obtain the basic field equations by varying the action. In section \ref%
{Sol}, we study charged black brane solutions with hyperscaling violation
in dilaton gravity in the presence of nonlinear electromagnetic
field. In Sec. \ref{Finite}, we study the finite action of the theory in
both the canonical and grand-canonical ensemble. The thermodynamics of
charged hyperscaling violating black branes will be investigated in Sec. \ref{Therm}.
We also generalize the Smarr formula of black branes to the case of
asymptotically Lifshitz charged black branes with hyperscaling violation.
In Sec. \ref{Stab}, we perform a thermal stability analysis of the
solutions. We finish our paper with some concluding remarks in section VI.


\section{FIELD EQUATION}

In this section we review the field equations of Einstein-dilaton gravity in
the presence of a linear and a nonlinear electromagnetic field. The gravity
part of the action in $(d+2)$-dimensions may be written as
\begin{equation}
I=\frac{1}{16\pi }\ \int_{\emph{M}}d^{d+2}x\sqrt{-g}R+\frac{1}{8\pi }\
\int_{\partial \emph{M}}d^{d+1}x\sqrt{-h}K+I_{M},  \label{Act1}
\end{equation}%
where $R$ is Ricci scalar, $\partial \emph{M}$ is the hypersurface at some
constant $r$, $h_{\alpha \beta }$ is the induced metric, $K$ is the trace of
the extrinsic curvature $K_{\mu \nu }=\nabla _{(\mu }n_{\nu )}$ of the
boundary and the unit vector $n^{\mu }$ is orthogonal to the boundary and
outward directed. The second term is the well-known Gibbons-Hawking term,
which is added to the action in order to have a well-defined variational
principle. The matter part of the action is
\begin{equation}
I_{M}=\frac{1}{16\pi }\int_{\emph{M}}d^{d+2}x\sqrt{-g}\ \left[ -\frac{1}{2}%
(\partial \Phi )^{2}+V(\Phi )-\frac{1}{4}e^{\lambda _{1}\Phi }H_{\mu \nu
}H^{\mu \nu }+\frac{1}{4}e^{\lambda _{2}\Phi }\left( -F\right) ^{s}\right] ,
\label{Im}
\end{equation}%
where $\lambda _{1}$ and $\lambda _{2}$ are free parameters of the model, $%
\Phi $ is the dilaton field, $F=F_{\mu \nu }F^{\mu \nu }$ is the Maxwel
invariant, $F_{\mu \nu }=\partial _{\lbrack \mu }A_{\nu ]}$ and $A_{\mu }$
is the nonlinear electromagnetic field. The linear electromagnetic field $%
H_{\mu \nu }=\partial _{\lbrack \mu }B_{\nu ]}$ with the Abelian gauge field
$B_{\nu }$ together with the dilaton field are to make the asymptotic of the
geometry to be the desired one. The nonlinear electromagnetic field is
required to have a nonlinear charged solution. For the potential of the
scalar field, motivated by the typical exponential potentials of string
theory, we will consider the following potential
\begin{equation}
V(\Phi )=-2\Lambda e^{\gamma \Phi },
\end{equation}%
where $\Lambda $ and $\gamma $ are two free parameters. The variation of the
total action $I=I_{G}+I_{M}$ with respect to gravitational field $g^{\mu \nu
}$, the scalar field $\Phi $ and the gauge fields $A_{\mu }$ and $B_{\mu }$
yields
\begin{eqnarray}
&&R_{\mu \nu }=\frac{1}{2}\partial _{\mu }\Phi \partial _{\nu }\Phi -\frac{%
V(\Phi )}{d}g_{\mu \nu }+\frac{1}{2}e^{\lambda _{1}\Phi }\left( H_{\;\mu
}^{\rho }H_{\;\rho \nu }-\frac{g_{\mu \nu }}{2d}H_{\lambda \rho }H^{\lambda
\rho }\right)  \notag \\
&&+\frac{1}{2}e^{\lambda _{2}\Phi }\left[ s\left( -F\right) ^{s-1}F_{\ \mu
}^{\rho }F_{\;\rho \nu }+\left( \frac{2s-1}{2d}\right) \left( -F\right)
^{s}g_{\mu \nu }\right]  \label{EMeq} \\
&&\nabla ^{2}\Phi =-\frac{dV(\Phi )}{d\Phi }+\frac{1}{4}\lambda
_{1}e^{\lambda _{1}\Phi }H_{\mu \nu }H^{\mu \nu }-\frac{1}{4}\lambda
_{2}e^{\lambda _{2}\Phi }\left( -F\right) ^{s},  \label{Phieq} \\
&&\partial _{\mu }\left( \sqrt{-g}e^{\lambda _{1}\Phi }H^{\mu \nu }\right)
=0,  \label{EMeq1} \\
&&\partial _{\mu }\left( s\sqrt{-g}e^{\lambda _{2}\Phi }\left( -F\right)
^{s-1}F^{\mu \nu }\right) =0.  \label{EMeq2}
\end{eqnarray}

\section{Black Brane Solutions with Hyperscaling Violation}

\label{Sol}

To find the nonlinear charged hyperscaling violating Lifshiz black branes, we will
closely follow the approach of \cite{Alishahiha:2012}. A suitable ansatz for
the metric, the dilaton field and the gauge fields of an isotropic static
spacetime may be written as
\begin{eqnarray}
ds^{2} &=&r^{2\alpha }\left( -r^{2z}f(r)dt^{2}+\frac{dr^{2}}{r^{2}f(r)}%
+r^{2}\sum\limits_{i=1}^{d}dx_{i}^{2}\right) ,\;\;\;\;  \label{ansatz} \\
\Phi &=&\Phi (r),\;\;\;\;\;F_{rt}=g(r),\ \ \ \ H_{rt}=h(r)\ ,
\end{eqnarray}%
while the other components of gauge fields are set to be zero. From the
Maxwell Eqs. (\ref{EMeq1}) and (\ref{EMeq2}), using the above ansatz, one
obtains
\begin{eqnarray}
H_{\;rt} &=&q_{1}e^{-\lambda _{1}\Phi }r^{\alpha (2-d)+z-d-1}, \\
F_{rt} &=&q_{2}e^{\frac{-\lambda _{2}\Phi }{2s-1}}r^{\frac{(2s-1)(z+2\alpha
-1)-d\alpha -d}{2s-1}}.
\end{eqnarray}%
Combining the $tt$ and $rr$ components of Eq. (\ref{EMeq}), one obtains
\begin{equation}
r^{2}(\partial _{r}\Phi )^{2}=2d(1+\alpha )(z+\alpha -1)\equiv \beta ^{2}.
\label{tr}
\end{equation}%
Thus, the scalar field is
\begin{equation}
\Phi (r)=\ln \left( \frac{r}{r_{0}}\right) ^{\beta }.  \label{Solphi}
\end{equation}%
Note that in order to have a real dilaton field, one has to assume $%
(1+\alpha )(z+\alpha -1)\geq 0$. Indeed, it can be seen that this assumption
is also the consequence of the null energy condition. More precisely consider a
null vector as $\xi ^{\mu }=(\sqrt{g^{rr}},\sqrt{g^{tt}},\mathbf{0})$, then
\begin{equation}
T_{\mu \nu }\xi ^{\mu }\xi ^{\nu }\varpropto d(1+\alpha )(z+\alpha
-1)r^{-2\alpha }f(r),
\end{equation}%
which is positive provided $(1+\alpha )(z+\alpha -1)\geq 0$. To find the
metric function $f(r)$ one may use any of the components of the field
equations (\ref{EMeq}). Using the $(ii)$-component of Eq. (\ref{EMeq})
and replacing $\Phi (r)$ from Eq. (\ref{Solphi}), we arrive at
\begin{eqnarray}
\left[ r^{d(\alpha +1)+z}f(r)\right] ^{\prime } &=&\frac{r^{\alpha
(d+2)+z+d-1}}{(\alpha +1)d}\left\{ -2\Lambda r_{0}^{-\gamma \beta }r^{\gamma
\beta }\right.  \notag \\
&&\left. -\frac{1}{2}r_{0}^{\lambda _{1}\beta }q_{1}^{2}r^{-2d(\alpha
+1)-\lambda _{1}\beta }-\frac{1}{4}(2s-1)(2q_{2}^{2})^{s}r_{0}^{\frac{%
\lambda _{2}\beta }{2s-1}}r^{\frac{-\lambda _{2}\beta -2sd(\alpha +1)}{2s-1}%
}\right\} .
\end{eqnarray}%
The above equation can be integrated to find the function $f(r)$ as
\begin{eqnarray*}
f(r) &=&-\frac{m}{r^{d+z+\alpha d}}+\frac{-2\Lambda r_{0}^{-\beta \gamma
}r^{\beta \gamma +2\alpha }}{d(\alpha +1)(\gamma \beta +\alpha (d+2)+z+d)} \\
&-&q_{1}^{2}r_{0}^{\lambda _{1}\beta }\frac{r^{-2\alpha (d-1)-\beta \lambda
_{1}-2d}}{2d(\alpha +1)[\alpha (2-d)+z-d-\beta \lambda _{1}]} \\
&-&\frac{1}{4d}\frac{(2q_{2}^{2})^{s}(2s-1)^{2}r_{0}^{\frac{\lambda
_{2}\beta }{2s-1}}r^{\frac{(-2+(4-2d)s)\alpha -2sd-\lambda _{2}\beta }{2s-1}}%
}{[(4s-d-2)\alpha -d-z+2sz-\lambda _{2}\beta ](\alpha +1)}.
\end{eqnarray*}%
where $m$ is the integration constant which is related to the mass of the
black brane as we will see later. In order to have the desired asymptotic
behavior for the metric, $f(r)$ should be equal to $1$ at infinity. Thus, we
should have
\begin{eqnarray*}
\gamma &=&-\frac{2\alpha }{\beta }, \\
\Lambda &=&-\frac{1}{2}r_{0}^{-2\alpha }(d(\alpha +1)+z-1)(d(\alpha +1)+z).
\end{eqnarray*}%
The function $f(r)$ should satisfy the equation of motion of the scalar
field. That is
\begin{eqnarray}
&&8\Lambda (\gamma d(\alpha +1)+\beta )r^{\gamma \beta }r_{0}^{-\gamma \beta
}+2(-\lambda _{1}d(\alpha +1)+\beta )r_{0}^{\beta \lambda _{1}}r^{-\beta
\lambda _{1}-2d(\alpha +1)}q_{1}^{2}  \notag \\
&&+(-\lambda _{2}d(\alpha +1)+\beta (2s-1))(2q_{2}^{2})^{s}r_{0}^{\frac{%
\beta \lambda _{2}}{2s-1}}r^{\frac{-\beta \lambda _{2}-2sd(\alpha +1)}{2s-1}%
}=0.
\end{eqnarray}%
This equation can be solved for the parameters $\lambda _{1},\lambda _{2}$
and $q_{1}$. One may find
\begin{eqnarray}
\lambda _{1} &=&-\frac{2\alpha (d-1)+2d}{\sqrt{2d(\alpha +1)(\alpha +z-1)}},
\notag \\
\lambda _{2} &=&(2s-1)\sqrt{\frac{2(\alpha +z-1)}{d(\alpha +1)}},  \notag \\
q_{1}^{2} &=&\frac{-4\Lambda (z-1)r_{0}^{2d(\alpha +1)}}{d(\alpha +1)+z-1}.
\end{eqnarray}%
Indeed, we have fixed the parameters $\Lambda $, $\gamma $, $\lambda _{1}$, $%
\lambda _{2}$ and $q_{1}$, while $q_{2}$ remains as an undetermined free
parameter which is related to the charge of the solution. It is, now,
important to check whether the other equations hold without imposing any
further constraints on the parameters of the solution. In particular one of
the nontrivial equations that need to be checked is the $tt$ component of
the Einstein equations of motion. Indeed, it is easy to see that this
equation is also satisfied without imposing any further constraints.

To summarize, the field equations admit the following nonlinear charged
black brane solution with hyperscaling violating factor. Note that in order
to follow the standard notation in the literature we set $\alpha =-\theta /d$%
, where $\theta$ is the hyperscaling violating exponent. It is also
notable to mention that these solutions are not valid for $\theta =d$ where $%
\alpha =-1$. Finally, we can rewrite the function $f(r)$ as%
\begin{equation}
f(r)=1-\frac{m}{r^{z+d-\theta }}+\frac{q^{2s}}{r^{\Gamma +d+z-\theta }},
\end{equation}%
where
\begin{eqnarray}
q^{2s} &=&\frac{(2s-1)r_{0}^{2(z-1-\frac{\theta }{d})}}{4(d-\theta )\Gamma }%
(2q_{2}^{2})^{s},  \label{qq} \\
\Gamma  &=&z-2+\frac{d-\theta }{2s-1}.  \label{Gam}
\end{eqnarray}%
It is worth mentioning that since\textbf{\ }$f(r)$\textbf{\ }should goes to $1$ at
infinity, we should have $z+d-\theta >0$ and $\Gamma +d+z-\theta >0$. But,
as we will show later $\Gamma >0$ and therefore $f(r)\rightarrow 1$ as $r$
goes to infinity provided $z+d-\theta >0$. The gauge and dilaton fields are
now given by
\begin{eqnarray}
H_{rt} &=&q_{1}r_{0}^{2(-\frac{\theta }{d}+\theta -d)}r^{(d-\theta +z-1)}, \\
F_{rt} &=&q_{2}r_{0}^{2(-\frac{\theta }{d}+z-1)}r^{-(\Gamma +1)}, \\[0.05in]
\Phi  &=&\ln \left( \frac{r}{r_{0}}\right) ^{\sqrt{2(d-\theta )(-\frac{%
\theta }{d}+z-1)}}.
\end{eqnarray}%
\begin{equation}
q_{1}^{2}=2(z-1)(z+d-\theta )r_{0}^{2(\frac{\theta }{d}+d-\theta )}.\label{q1}
\end{equation}%
Equation (\ref{q1}) shows that since $z+d-\theta >0$, $z$ should be larger than $1$
too.
\section{FINITE ACTION IN CANONICAL AND GRAND-CANONICAL ENSEMBLES \label%
{Finite}}
In general, the total action $I$ given in Eq. (\ref{Act1}) is divergent when
evaluated on a solution. One way of dealing with the divergences of the
action is adding some counterterms to the action (\ref{Act1}). The
counterterms should contain a part which removes the divergence of the
gravity part of the action and a part for dealing with the divergence of the
matter action. Since the horizon of our solution is flat, the counterterm
which removes the divergence of the gravity part should be proportional to $%
r^{\theta/d}\sqrt{h}$. The counterterm for the matter part of the action in
the absence of dilaton ($\lambda _{1}=0$) and for the case of Lifshitz
solution ($\theta =0$) has been introduced in Ref. \cite{saremi2009}. Here,
we generalize it to the case of Lifshitz solutions with hyperscaling violating factor
in the presence of dilaton field. For this case, due to the fact that on the
boundary $e^{\lambda _{1}\Phi \left( r\right) }B_{\gamma }B^{\gamma }$ is
constant for our solutions, the following counterterms make the action
finite: \textbf{\ }
\begin{equation*}
I_{ct}=-\frac{1}{16\pi }\int_{\partial M}\ d^{d+1}x\ \sqrt{-h}r^{\theta /d}\left[
2(d-\theta )-a\left( -e^{\lambda _{1}\Phi}B_{\gamma }B^{\gamma }\right) ^{1/2}\right] ,
\end{equation*}%
where $a=\sqrt[~]{2\left( z-1\right) \left( d+z-\theta \right) }$. Note that
because of the constraints on $d$, $z$ and $\theta $ explained in the last
section, $a$ is real as it should be. The variation of the total action $%
\left( I_{tot}\ =I+I_{ct}\right) $ about the solutions of the equations of
motion is
\begin{equation}
\delta I_{tot}=\int d^{d+1}x\left( S_{\alpha \beta }\delta h^{\alpha \beta
}+S_{\alpha }^{L}\delta B^{\alpha }\right) +\frac{1}{16\pi }\int d^{d+1}x%
\sqrt{-h}s(-F)^{s-1}e^{\lambda _{2}\Phi }n^{\mu }F_{\ \mu \alpha
}\delta A^{\alpha },  \label{Ivar}
\end{equation}%
where
\begin{equation}
S_{\alpha \beta }=\frac{\sqrt{-h}}{16\pi }\left\{ \Pi _{\alpha \beta
}+r^{\theta /d}\left[(d-\theta )h_{\alpha \beta }-\frac{a}{2} e^{\lambda _{1}\Phi/2}
\left( -B_{\gamma }B^{\gamma }\right) ^{-1/2}\left( B_{\alpha }B_{\beta
}-B_{\gamma }B^{\gamma }h_{\alpha \beta }\right)\right] \right\} ,
\end{equation}%
\begin{equation}
S_{\beta }^{L}=-\frac{\sqrt{-h}}{16\pi }\left\{ n^{\alpha }H_{\alpha
\beta }+a e^{\lambda _{1}\Phi/2}\left(
-B_{\gamma }B^{\gamma }\right) ^{-1/2}B_{\ \beta }\right\} ,
\end{equation}%
with $\Pi _{\alpha \beta }=K_{\alpha \beta }-Kh_{\alpha \beta }$.

Equation (\ref{Ivar}) shows that the variation of the total action with
respect to $A^{\mu }$ will only give the equation of motion of the nonlinear
massless field $A^{\mu }$ provided the variation is at fixed nonlinear
massless gauge potential on the boundary. Thus, the total action, $I_{tot}\
=I+I_{ct}$, given in Eqs. \ref{Ivar} is appropriate for the grand-canonical
ensemble, where $\delta A^{\mu }=0$ on the boundary. But in the canonical
ensemble, where the electric charge $s(-F)^{s-1}e^{\lambda _{2}\Phi
}n^{\mu }F_{\ \mu \alpha }A^{\alpha }$\ is fixed on the boundary, the
appropriate action is
\begin{equation}
I_{tot}=I+I_{ct}-\frac{1}{16\pi }\int_{\partial M}d^{d+1}x\ \sqrt[\ ]{-h}\
se^{\lambda _{2}\Phi }n^{\mu }(-F)^{s-1}F_{\ \mu \alpha }A^{\alpha
}.  \label{I3term}
\end{equation}%
The last term in Eq. (\ref{I3term}) is the generalization of the boundary
term introduced by Hawking for linear electromagnetic field  \cite%
{S. W. Hawking grand} and in Ref. \cite{DSV} for nonlinear gauge field.
Thus, both in canonical and grand-canonical ensemble, the
variation of total action about the solutions of the field equations is
\begin{equation}
\delta I_{tot}=\int d^{d+1}x\left( S_{\alpha \beta }\delta h_{\alpha \beta
}+S_{\alpha }^{L}\delta B^{\alpha }\right) .
\end{equation}%
That is, the nonlinear gauge field is absent in the variation of the total
action both in canonical and grand-canonical ensembles, and therefore, as in
the absence of massless electromagnetic field, the dual field theory for
a hyperscaling violating Lifshitz  spacetime in the presence of a nonlinear
electromagnetic field has a stress tensor complex consisting of the energy
density $\varepsilon $, energy flux $\varepsilon _{i}$, momentum density $%
P_{i}$ and spatial stress tensor $P_{ij}$ satisfying the conservation
equations
\begin{eqnarray}
&& \partial _{t}\varepsilon +\partial _{i}\varepsilon ^{i}=0,\ \ \ \ \ \ \ \ \ \ \
\ \ \ \  \partial _{t}P_{j}+\partial _{i}P_{\ j}^{i}=0,\\
&& \varepsilon =2S_{\ t}^{t}-S_{L}^{t}B_t,\ \ \ \ \ \ \ \ \ \ \ \ \ \varepsilon
^{i}=2S_{\ t}^{i}-S_{L}^{i}B_t,  \label{energy-density}\\
&& P_{i}=-2S_{\ i}^{t}+S_{L}^{t}B_i,\ \ \ \ \ \ \ \ \ P_{\ i}^{j}=-2S_{\
i}^{j}+S_{L}^{j}B_i.
\end{eqnarray}

\section{THERMODYNAMICS OF NONLINEAR CHARGED HYPERSCALING VIOLATING BLACK BRANES \label{Therm}}

Now, we investigate thermodynamics of charged Lifshitz black branes with
hyperscaling violating factor. One can obtain the temperature of the event horizon by
using
\begin{equation}
T=\frac{1}{2\pi }\left( -\frac{1}{2}\nabla _{b}\zeta _{a}\nabla ^{b}\zeta
^{a}\right) _{r=r_{+}}^{1/2}.
\end{equation}%
Using the fact that the mass parameter is
\begin{equation}
m=r_{+}^{z-\theta +d}+q^{2s}r_{+}^{-\Gamma },  \label{mass}
\end{equation}%
one obtains
\begin{equation}
T=\frac{1}{4\pi }\left\{ (d+z-\theta )r_{+}^{z}-\Gamma q^{2s}r_{+}^{-(\Gamma
+d-\theta )}\right\} .  \label{Temp}
\end{equation}%
One may note that there exists an extreme black hole. The charge and mass of
the extreme black hole is%
\begin{eqnarray*}
q_{ext}^{2s} &=&\frac{z+d-\theta }{\Gamma }r_{ext}^{z+d-\theta +\Gamma }, \\
m_{ext} &=&\frac{\Gamma +z+d-\theta }{\Gamma }r_{ext}^{z+d-\theta },
\end{eqnarray*}%
and therefore the condition of having black holes is%
\begin{equation}
\left( \frac{\Gamma m}{\Gamma +z+d-\theta }\right) ^{\Gamma +z+d-\theta
}\geq \left( \frac{\Gamma q^{2s}}{z+d-\theta }\right) ^{z+d-\theta }.
\end{equation}%
The entropy per unit volume of the horizon is given, as usual, by the
Bekenstein-Hawking formula
\begin{equation}
S=\frac{1}{4}r_{+}^{d-\theta }\ .
\end{equation}%
Using Eq. (\ref{energy-density}) one obtains the energy density of black
brane as
\begin{equation}
\varepsilon =\frac{d-\theta }{16\pi }m,
\end{equation}%
where $m$ in terms of $r_{+}$ and $q$ is given in Eq. (\ref{mass}).

Before starting the calculations of the electric charge and potential, one
should note that the linear gauge field $B_{\mu }$ is just needed to support
the structure of the asymptotic of the Lifshitz spacetime solution with
hyperscaling violating factor. In other words, it does not have a thermodynamic
interpretation. The fact that this gauge field does not affect the
thermodynamics is due to the fact that its charge parameter is not a free
parameter and it is completely determined by the parameters of the metric.
The electric charge density of the gauge field $A_{\mu }$ may be calculated
by using%
\begin{equation*}
Q=\frac{1}{16\pi \Omega }\int d\Omega r^{d-\theta }e^{\lambda _{2}\Phi
}\left( -F\right) ^{s-1}F_{\mu \nu }n^{\mu }u^{\nu },
\end{equation*}%
where $n^{\mu }$ and $u^{\nu }$are the spacelike and timelike unit normals
to a sphere of radius $r$,%
\begin{equation*}
u^{\nu }=\frac{1}{\sqrt{-g_{tt}}}dt=\frac{1}{r^{z}\sqrt{f}}dt,%
n^{\mu }=\frac{1}{\sqrt{g_{rr}}}dr=r\sqrt{f}dr.
\end{equation*}%
One obtains
\begin{equation}
Q=\frac{s}{16\pi }2^{s-1}(q_{2})^{2s-1}.
\end{equation}%
Choosing the infinity as the reference point of the potential, the nonlinear
gauge field $A_{t}$ can be obtained as%
\begin{equation*}
A_{t}={\int^r}_{\infty} F_{rt}dr=-\frac{1}{\Gamma }r_{0}^{2(-\frac{\theta }{d}+z-1)}\frac{%
q_{2}}{r^{\Gamma }},
\end{equation*}%
where $\Gamma $ is given in Eq. (\ref{Gam}). To avoid the divergence of $%
A_{t}$, we must have $\Gamma >0$. Now, using the definition of electric
potential at infinity with respect to the horizon
\begin{equation}
U=A_{\mu }\chi ^{\mu }\left\vert _{r\rightarrow \infty }-A_{\mu }\chi ^{\mu
}\right\vert _{r=r_{+}},
\end{equation}%
where $\chi ^{\mu }=\partial /\partial t$ is the null generators of the
event horizon, the electric potential is obtained as
\begin{equation}
U=\frac{q_{2}}{\Gamma }r_{0}^{2(-\frac{\theta }{d}+z-1)}r_{+}^{-\Gamma }.
\label{Pot}
\end{equation}%
Now, it is a matter of straightforward calculation to show that the first
law of thermodynamics holds on the black brane horizon:
\begin{equation}
d\varepsilon =TdS+UdQ\ .
\end{equation}%
Specifically, the temperature and potential given in Eqs. (\ref{Temp}) and (%
\ref{Pot}) can be reobtained by the following equations:
\begin{equation}
T=\left( \frac{\partial \varepsilon }{\partial S}\right) _{Q}\ ,\quad
U=\left( \frac{\partial \varepsilon }{\partial Q}\right) _{S}\,.
\end{equation}%
The Smarr formula can be obtained as
\begin{equation}
\varepsilon =\frac{\left( d-\theta \right) }{\left( d-\theta +z\right) }%
\left[ TS+\frac{(\Gamma +z+d-\theta )(2s-1)}{2s(d-\theta )}Q\Phi \right] .
\end{equation}%

\section{STABILITY OF NONLINEAR CHARGED BLACK BRANES WITH HYPERSCALING VIOLATION\label{Stab}}

Finally, we investigate the stability of charged black brane solutions of
Einstein nonlinear Maxwell gravity both in canonical and grand-canonical
ensemble. The local stability can in principle be carried out by finding the
determinant of the Hessian matrix of $\varepsilon (S,Q)$ with respect to its
extensive variables $S$ and $Q$ ($H_{X_{i}X_{j}}^{\varepsilon }=[\partial
^{2}\varepsilon /\partial X_{i}\partial X_{j}]$) \cite{HESS}. The number of
thermodynamic variables depends on the ensemble that is used. In the
canonical ensemble, the charge is a fixed parameter, and therefore the
positivity of the heat capacity $C_{Q}=T_{+}/(\partial ^{2}\varepsilon
/\partial S^{2})_{Q}$ is sufficient to ensure the local stability. The heat
capacity can be obtained as
\begin{eqnarray}
C_{Q} &=&T\left( \frac{\partial S}{\partial T}\right) _{Q}=\left( \frac{%
\partial \varepsilon /\partial r_{+}}{\partial T/\partial r_{+}}\right) _{Q} \notag
\\
&=&\frac{(d-\theta )Tr_{+}^{d-\theta }}{z(z+d-\theta )r_{+}^{z}+\Gamma
\lbrack \Gamma +d-\theta ]q^{2s}r_{+}^{-[d-\theta +\Gamma ]}}.
\end{eqnarray}%
Since as we mentioned before both $z+d-\theta $ and $\Gamma $ are positive,
we conclude that the specific heat at constant charge is always positive
provided $d>\theta $. That is, the black brane solutions are stable in
canonical ensemble for $d>\theta$.

In the grand-canonical ensemble, the determinant of the Hessian matrix of
the energy with respect to $S$ and $Q$ can be obtained as%
\begin{eqnarray}
H_{s,Q}^{\varepsilon } &=&\frac{64r^{-(\Gamma +d-\theta)}}{%
s(2s-1)^{1/s}q^{2(s-1)}}\left\{ \frac{4(d-\theta )\Gamma }{%
r_{0}^{2(z-1-\theta /d)}}\right\} ^{-\frac{2s-1}{s}}\digamma  \notag \\
\digamma &=&z(z+d-\theta )r_{+}^{z}-\Gamma
(2s-1)(z-2)q^{2s}r_{+}^{-\left(\Gamma+d-\theta\right) }  \notag \\
&=&4\pi zT+2\Gamma[(z-1)+s(2-z)]q^{2s}r_{+}^{-\left(\Gamma+d-\theta\right)}.  \label{Hes}
\end{eqnarray}%
Assuming $d>\theta $ and $s>1/2$, one can easily see from Eq. (\ref{Hes})
that the only factor in the Hessian matrix which can be negative is $%
\digamma $.

First, we investigate the condition of having a stable black hole. In the
case of linear Maxwell field ($s=1$), $\digamma >0$ and therefore the linear
charged Lifshitz black branes with hyperscaling violating factor is always stable in
grand-canonical ensemble. In the case of nonlinear electromagnetic field ($%
s>1/2$), $\digamma $ is positive for $1<z\leq 2$ and therefore the black
hole solution with $1<z\leq 2$ is stable in grand-canonical ensemble. Of
course, one may remember that $\Gamma $ should be positive and therefore $%
d-\theta >(2s-1)(2-z)$. However, for the case of $z>2$, $\digamma $ is
positive provided:%
\begin{equation*}
s\leq \frac{z-1}{z-2},
\end{equation*}%
or%
\begin{equation*}
z\leq \frac{2s-1}{s-1}.
\end{equation*}%
It is worth mentioning that in this case since $z$ is larger
than two, $s$ should be larger than $1$.

\begin{figure}[tbp]
{\scriptsize \center
\includegraphics[scale=.5]{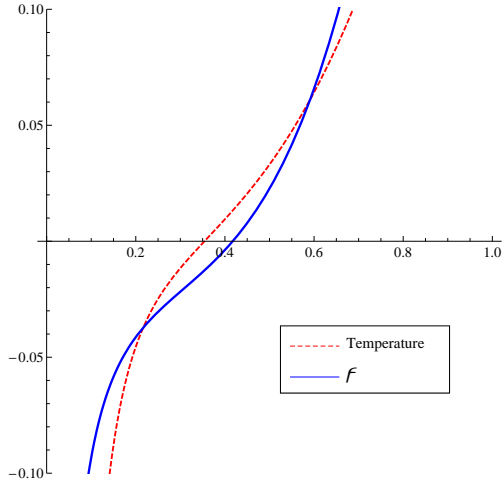} }
\caption{{\protect\small {Temperature (red,dotted) and $\digamma$ (blue,line) as
functions of the horizon radius for $q=0.6$, $d=3$, $z=3$ $s=4$ and $\theta=2$.}}}
\label{Fig1}
\end{figure}

Second, we investigate the unstable phase of nonlinear charged hyperscaling violating Lifshitz black branes.
Indeed, the black brane solution is unstable
provided%
\begin{equation*}
z>\frac{2s-1}{s-1},
\end{equation*}%
or%
\begin{equation*}
s>\frac{z-1}{z-2}.
\end{equation*}%
Note that in either of the cases $s$ is larger than $1$. To be more clear,
we plot a case which can have an unstable phase. As one can see in Fig \ref%
{Fig1}, the small black branes $r_{ext}\leq r_{+}<r_{+\max }$ is unstable,
while the large black branes are stable. Note that in Fig. \ref{Fig1} the black
branes with horizon radius in the range (0.354-0.418) are unstable.

\section{Concluding remarks}

In this paper, we considered asymptotically Lifshitz black branes with
hyperscaling violation in the presence of a massless nonlinear gauge
field in dilaton gravity. Indeed, there are two massless electromagnetic
fields coupled with the dilaton field. The first one which is linear
accompanied with the dilaton field makes the desired asymptotic for the
solution, while the second nonlinear gauge field gives charge to the
solution. After presenting an analytic nonlinear charged hyperscaling violating Lifshitz black
brane solution, we present the action in
both the canonical and grand-canonical ensemble. Indeed, the appropriate
action of a charged black hole in the grand-canonical and canonical
ensembles are not the same. The appropriate actions in these ensembles in
the absence of a dilaton field are given by Hawking and Ross for a linear
electromagnetic field \cite{S. W. Hawking grand}. Its generalization to the
case of nonlinear gauge field in the absence of dilaton field has been
introduced in \cite{DSV}. Here, we generalized this action to the case of
power-law electromagnetic theory in the presence of dilaton field and
introduced the appropriate action for both the canonical and grand-canonical
ensembles. Next, we generalized the counterterm method introduced in Ref.
\cite{saremi2009} to our case and used it to calculate the finite energy
density. We also found the general thermodynamic relationship for the energy
density in terms of the extensive thermodynamic quantities, entropy, and
charge density, and their intensive conjugate quantities, temperature and
electric potential. This result generalizes the well-known Smarr formula of
AdS black holes and reduces to the Smarr formula of the 4-dimensional uncharged
Lifshitz solution of Ref. \cite{Peet1}. Finally, we investigated the stability of the
solutions in both the canonical and grand-canonical ensembles. Before,
summarizing the results, it is worth mentioning the constraint on the
parameters $z$, $\theta $, $s$ and $d$. First, since the electric potential
should vanish at infinity $d-\theta +(2s-1)(z-2)>0$. Second, since $f(r)$
should go to one as $r$ goes to infinity, $z+d-\theta >0$. Finally, since $%
q_{1}$ should be real, $z>1$. We found that for $d>\theta $ while the
solution is thermally stable in canonical ensemble, it can has an unstable
phase in grand-canonical ensemble. Indeed, this instability is the effect of
the nonlinearity of the electromagnetic field. We found that when the field
is linear ($s=1$), the black brane solution is stable. We also found that
for $z\leq 2$, the solution is stable even in grand-canonical ensemble.

The thermodynamics of rotating Lifshitz black branes in the presence of
nonlinear massless  gauge field has been investigated in \cite{Rot}. In Ref.
\cite{Rot}, the counterterm method has  been generalized to the case of
rotating Lifshitz solution. Here, we applied the counterterm method only to
the case of static solutions. It would be interesting to generalize this
method to the case of rotating black branes. Further work in this area will
involve considering the thermodynamics of Lifshitz black branes of modified
theory of gravity in the presence of a dilaton field with or without
hyperscaling violating factor. These works are in progress.

\acknowledgments{We thank Shiraz University Research Council. This
work has been supported financially by Research Institute for
Astronomy and Astrophysics of Maragha, Iran.}


\end{document}